\documentclass{article}
\usepackage{fullpage}
\usepackage{amsmath}
\usepackage{times}
\usepackage{graphicx} 
\usepackage{subfig}
\graphicspath{{./figs/}}

\usepackage{hyperref}

\usepackage{authblk}

\title{Deep rank-based transposition-invariant distances on musical sequences}

\author[1, 2]{Ga\"etan Hadjeres}
\author[3, 4]{Frank Nielsen}

 \affil[1]{LIP6, Universit\'e Pierre et Marie Curie}
 \affil[2]{Sony CSL, Paris}
 \affil[3]{\'Ecole Polytechnique, Palaiseau, France}
 \affil[4]{Sony CSL, Tokyo}

\date{}
\begin{document}

\maketitle

\expandafter\def\expandafter\UrlBreaks\expandafter{\UrlBreaks
  \do\-}

\begin{abstract}
  Distances on symbolic musical sequences are needed for a variety of
  applications, from music retrieval to automatic music generation. These
  musical sequences belong to a given corpus (or style) and it is obvious that a
  good distance on musical sequences should take this information into account;
  being able to define a distance ex nihilo which could be applicable to all
  music styles seems implausible. A distance could also be invariant under some
  transformations, such as transpositions, so that it can be used as a distance
  between musical motives rather than musical sequences.  However, to our
  knowledge, none of the approaches to devise musical distances seem to address
  these issues.  This paper introduces a method to build transposition-invariant
  distances on symbolic musical sequences which are learned from data. It is a
  hybrid distance which combines learned feature representations of musical
  sequences with a handcrafted rank distance. This distance depends less on the
  musical encoding of the data than previous methods and gives perceptually good results.
  We demonstrate its efficiency on the dataset of \emph{chorale melodies by J.S. Bach}.
\end{abstract}

\section{Introduction}
\label{sec:introduction}
Determining whether two musical sequences are similar or not is a key ingredient in music
composition. Indeed, the repeated occurrences of a given pattern (transformed or
not) is easily perceived by an attentive listener. Among possible
transformations of a pattern, we can cite
\begin{itemize}
\item transposition
\item mode change
\item augmentation / diminution  
\item ornamentation / simplification.
\end{itemize}

These patterns in music gives the listener expectations of what could
follow. This latter is then gratified to have guessed right or can be
surprised by a pleasing or unexpected change. The musical pieces may
then appear more coherent and organized. It is then up to the composer
to play with this series of fulfilled/unfilled expectations. From a
compositional point of view, this allows a composer to create long
pieces of music while retaining the listener's attention.

\begin{figure*}
  \centering
      \includegraphics[width=0.8 \textwidth]{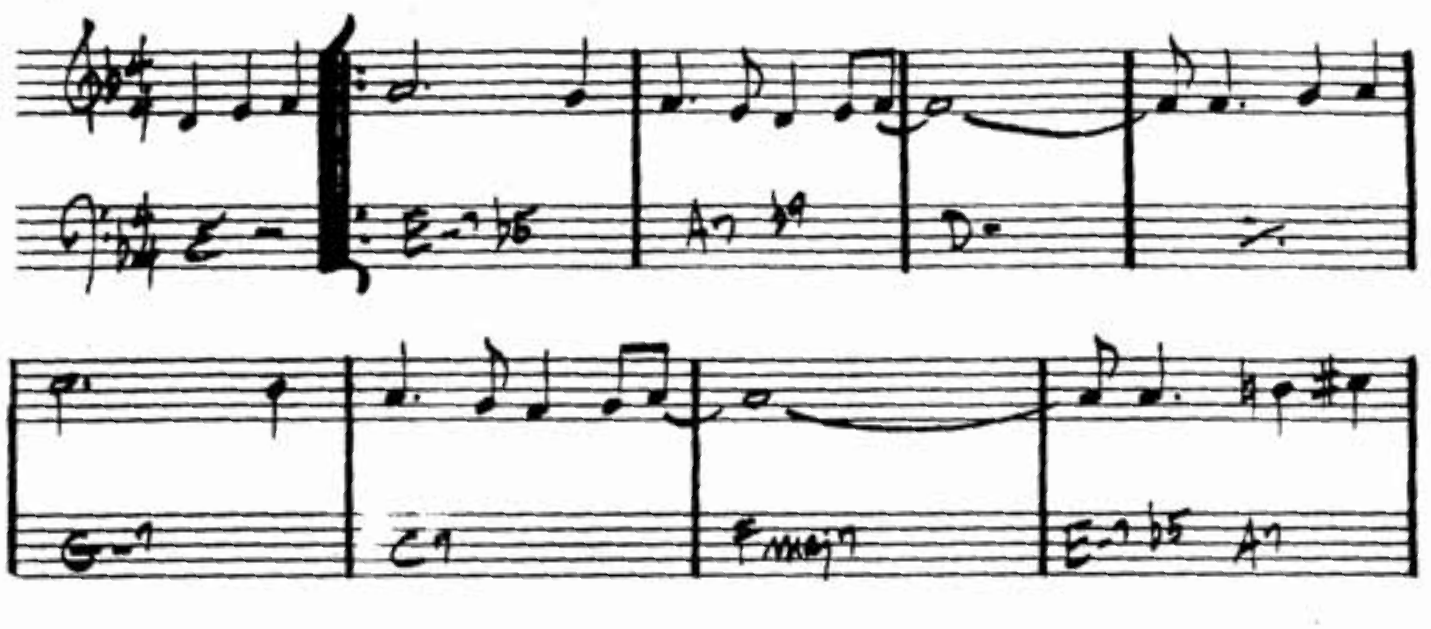}
      \caption{First two staves of the jazz standard \emph{Beautiful
          Love}. The first four measures are transposed a minor third
        higher, adapted to the mode of F major (instead of D minor)
        and the rhythm is slightly changed.}
  \label{fig:beautifulLove}
\end{figure*}

Many musical pieces are intrinsically-based on the different
repetitions of a given pattern. Fugues or sonatas are examples of
such pieces where the overall structure results from how the
occurrences and transformations of a given pattern unfold through
time. But this is also true on a local scale in many musical genres
since patterns give coherence between musical phrases. This is
particularly observable in pop and jazz songs (see
Fig.~\ref{fig:beautifulLove} for instance for an example of a pattern
and its transformation).

Due to the omnipresence of patterns in music, good distances on
musical sequences are essential and can be used for a wide variety of
tasks:
\begin{itemize}
\item plagiarism detection \cite{cason2012singing}
\item music retrieval \cite{Allali2010}
\item automatic musical analysis \cite{david2014vers,giraud2015computational}
\item automatic music generation \cite{DBLP:journals/corr/RoyPP17}.
\end{itemize}

The traditional distance used for sequence comparison is the
edit-distance (also known as the Levenshtein distance
\cite{levenshtein1966binary}).  On sequences of symbols, it basically
consists in assigning a cost to different basic edit operations such
as the insertion, the deletion or the substitution of a single
symbol. The distance between two sequences is then the cost of the
minimal sequences of basic edit operations allowing to change the
first sequence into the second one. This minimal cost is easily
calculated using a dynamic programming algorithm.

Most of the existing distances on musical sequences are based on
generalizations of the aforementioned distance by taking into account
the specificity of the set of musical sequences (the edit distance was
primarily designed for text).

For instance, \cite{mongeau1990comparison} propose to extend the set
of basic edit operations by two other operations that are more
specific to musical sequences: fragmentation and consolidation.

The main issue with this approach is that the edit-distance strongly
depends on the choice of the encoding of the musical content. Indeed,
contrary to textual data, musical content (such as monophonic
melodies) can be encoded in \emph{many ways} and there are \emph{a priori}
no representation which is better than another.

The importance of the choice of the data representation is pinpointed
in \cite{cambouropoulos2001pattern}. In this paper, the authors argue
that the MIDI pitch representation is insufficient for applications in
tonal music as it disregards the hierarchical importance of diatonic
scale tones over the 12-tone discrete pitch space. To address this
issue, they propose a new representation, called the General Pitch
Interval Representation. It is a representation that takes into
account the diatonic intervals in scale steps and other more abstract
representations such as contour strings (the contour string of a
melody is a representation where only the following events are
considered: repeat, ascending or descending step, ascending or
descending leap).

The interest in such a representation is that it introduces
perceptually salient information directly into the sequence encoding.

This idea is further explored in \cite{grachten2004melodic} where they
propose to encode a melody using its Implication/Realization
structure. It is a concept drawn from the theory of perception and
cognition of melodies from \cite{narmour1992analysis} which is based
on the Gestalt theory. It consists in assigning different basic
structures depending on the contour of a sequence and can be
considered as a generalization of the contour string encoding.

Using contour strings encoding implies that a melody is only considered
up to a transposition since only intervals with respect to the
previous note is considered \cite{lemstrom2000including}. The same
idea can also be used by considering ratios of rhythmic patterns
instead of their absolute values.  We refer the reader to
\cite{hanna2007optimizing} where the authors study the advantages and
drawbacks of many monophonic sequence representations on the
edit-distance algorithms.

Geometric interpretations of the distance between two musical
sequences have also been proposed
\cite{doi:10.1076/jnmr.31.4.321.14162,toussaint2003algorithmic}. Their
advantage is to be applicable on polyphonic sequences contrary to the
methods based on the
edit-distance. \cite{doi:10.1177/102986490701100106} compares the two
approaches on a variety of music retrieval tasks.

However, defining how close two sequences are is in fact an
ill-defined problem. This notion is very \emph{subjective} and it seems
implausible to find a universal rule applying to every musical style
and sequences: sequences can be ``close'' with respect to a given
music style and ``far'' in another music style. Furthermore, attempts
to ground it on a psychological basis using an appropriate
representation still suffer from the arbitrariness of the underlying
distance.

In this work, we introduce a corpus-dependent distance between
two musical sequences.  Our distance relies on a permutation-based
distance \cite{amato2016deep} applied on high-level features obtained
via a sequence-to-sequence autoencoder.  This approach is general and
we believe more independent of the choice of the representation than
all previous methods, and can be applied to both monophonic and
polyphonic music.

We then extend our method in order to obtain a way to generate
transposition-invariant distances, which means that a sequence and its
transposition should be close. Contrary to other methods, this
distance is made transposition-invariant without changing the sequence
encoding.

In the following, we focus on monophonic sequences with a given
representation, but these ideas can easily be generalized to other
representations of musical material, from monophonic to polyphonic
ones.


Our contributions are the following:
\begin{itemize}
\item introduction of a framework to build distances on sequences,
\item extension of this approach in order to construct invariant
  distances with respect to a given set of transformations,
\item introduction of corpus-dependent musical distances in music.
\end{itemize}

We believe that linking the distance between musical sequences to the
specific genre of these musical sequences is a way to address issues
related to the choice of a perceptually-appropriate distance and is
more likely to yield better results.

The outline of this paper is the following: In
Sect.~\ref{sec:related-works}, we expose related works about
transformation-invariant features and transformation-invariant
distances; Section~\ref{sec:model} introduce our method to construct
a data-dependent distance on sequences and Sect.~\ref{sec:tid}
improves this method in order to obtain a distance which is invariant
with respect to a set of transformations; Finally, in
Sect.~\ref{sec:res} we present experimental results about the
introduced distances and highlight their efficiency on the dataset of
the \emph{chorale melodies from the J.S. Bach chorale harmonizations.}

\section{Related works}
\label{sec:related-works}
Finding transformation-invariant distances is a problem which was primarily
addressed on image datasets. Indeed, learning distances from a corpus of images
is crucial in many applications (classification, clustering, or retrieval tasks) and it is
often desirable that this distance be independent under some natural
transformations on images such as rotations and translations.

Taking into account the ``natural'' set of transformations which acts on a
dataset is of a great interest since we can use this information to obtain more
robust and more informative feature representations.

The feature representations
need not necessarily be invariant with respect to a set of transformations, but
sometimes only equivariant \cite{cohen2016group}. Equivariance here means that
the feature representation of a transformed image can be obtained by applying a
known transformation directly to the feature representation of the original
image.
      
Convolutional Neural Networks (CNNs)
\cite{krizhevsky2012imagenet,simonyan2014very} for instance take into account
the importance of translations on image datasets by using the same transposed
filter. A generalization of the regular CNN filters has been proposed
\cite{DBLP:journals/corr/WorrallGTB16}. It aims at obtaining a CNN which is
equivariant to both translations and rotations. It is in fact possible to devise
more general approaches which are able to deal with any symmetry group acting on
images as shown in
\cite{NIPS2014_5424,cohen2016steerable,2017arXiv170603078B}. In
\cite{2017arXiv170603078B}, the proposed model is suitable for a theoretical
analysis about the stability of the learned invariant representations.
      
However, equivariant feature representations are not particularly
suitable for building invariant feature representations. Indeed, in
order to obtain an invariant representation, one would have to average
all feature representations of all possible image transformations
under the chosen group, which is either intractable or computationally
demanding.

An approach in the context of shape matching which shares the same motivations
as ours can be found in \cite{manay2006integral}. Contrary to images, two shapes
are considered to be identical if we can obtain one by applying a group
transformation on the other one. These group transformations can be as above the
group of displacements (translations and rotations), but can also be, in the
context of shapes, the affine group (translations, rotations and rescalings). A
distance between shapes is then constructed by introducing a distance on the
integral invariants of a shape. These handcrafted quantities are invariant with
respect to the group of transformations acting on shape contours and thus assert
that the constructed distance is well-defined on shapes. The main difference
with the approaches in image is that the feature representations (here the
integral invariants) is not learned from data but constructed by hand.

In comparison, our method is able to learn transformation-invariant feature
representations from data resulting in a transformation-invariant distance.

\section{Corpus-based distance}
\label{sec:model}

\begin{figure}

  \centering
    \subfloat[]{
      \includegraphics[height=0.9cm]{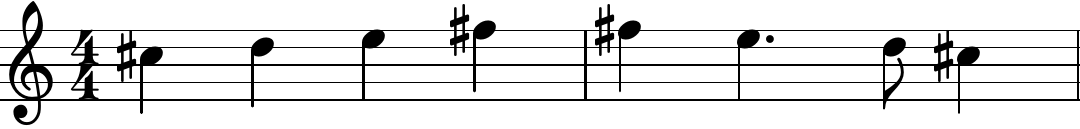}
      \label{fig:seq}

      }
\hfill
  \subfloat[]{
    \includegraphics[height=1cm, trim=0 0 0 3, clip=true]{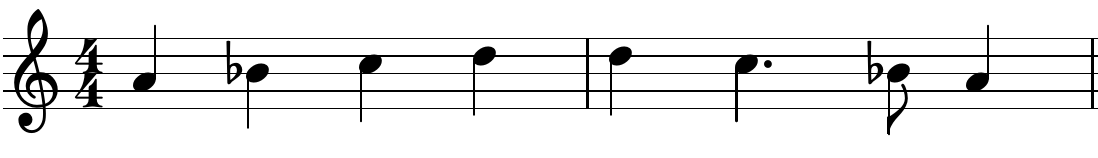}
     \label{fig:seqtransposed}
    
}

  \caption[BWV434 caption]{(a) a 2-beat sequence and (b) its transposition a major third lower.}
  \label{fig:transpoexample}
\end{figure}
\subsection{Rank-based distance}
\label{sec:pbd}
Our method to construct distances on sequences relies on the
rank-based (or permutation-based) distances described in \cite{amato2016deep}. The
idea is to define a distance based solely on the ranking of the
high-level activations of a Deep Neural Network (DNN)
\cite{Goodfellow-et-al-2016}. Indeed, activations of neurons from
high-level hidden layers can be used as features for describing the
input. Each of this feature can encode a particular concept about the
input and how these concepts are ranked with respect to one another is
sufficient to determine if two inputs are similar.

More precisely, let $x$ be a $N$-dimensional feature vector. We define
$\Pi(x)$ to be the vector of ranks of $x$. This vector is a
permutation of the $N$-tuple $[N]$ such that
\begin{equation}
  \label{eq:2}
 \forall i, j \in [N], \quad i < j \implies x_{\Pi(x)_i} \geq x_{\Pi(x)_j}.
\end{equation}

With this notation, $\Pi(x)_1$ is the index $i$ such that $x_i$ is the
greatest coordinate of $x$.

Given two feature vectors $x$ and $y$, we can define their distance to
be the distance between their permutations $\Pi(x)$ and $\Pi(y)$ using
popular distances between permutations. In the following, we will
consider two popular rank correlation measures: Spearman's rho
distance and Kendall's tau distance.

In its simplest formulation,  Spearman's rho distance $\rho$ between feature vectors $x$ and $y$ is defined to be the $\ell_2$ norm between their vectors of ranks. This gives us:
\begin{equation}
  \label{eq:1}
 \rho(x, y) = \sqrt{\sum_{i = 1}^N (\Pi(x)_i - \Pi(y)_i)^2}.
\end{equation}

The Kendall tau distance  is a measure of similarity between two rank variables. It is based on the number of pairwise inversions needed to change one ordering into the other.
Given two vectors of ranks $\Pi(x), \Pi(y)$ of size $N$, we say that a pair of integers $i < j$  is concordant if $\Pi(x)_i < \Pi(x)_j$ and $\Pi(y)_i < \Pi(y)_j$ and discordant if $\Pi(x)_i < \Pi(x)_j$ and $\Pi(y)_i > \Pi(y)_j$ (and these definitions also hold when we reverse all inequalities). Since there are $N(N-1)/2$ such pairs, we can define a similarity on feature vectors $x$ and $y$ by
\begin{equation}
  \label{eq:9}
  \tau(x, y) = \frac{ \sharp\left\{\textrm{concordant pairs}\right\} - \sharp\left\{\textrm{discordant pairs}\right\}}{N(N-1)/2},
\end{equation}
where $\sharp\left\{\textrm{concordant pairs}\right\}$ indicates the number of concordant pairs when considering the rank vectors of $x$ and $y$.
This similarity measure is in $[-1, 1]$ and equals $1$ when the ranks are equal.

It is worth noting that computing  Kendall's tau distance is more computationally-demanding than computing Spearman's rho distance due to the quadratic number of terms.

\subsection{Sequence-to-Sequence autoencoder}
\label{sec:s2sae}
In the following, we consider that we have a i.i.d. dataset
$\mathcal{D}$ of $K$ sequences of length $L$
\begin{equation}
  \label{eq:5}  
  \mathcal{D} := \left\{s^{(k)} = (s_1^{(k)}, \dots, s_L^{(k)}), \quad s_i \in [A]  \right\}_{k \in [K]},
\end{equation}
where sequences are composed of tokens in $[A]$ with
$A$ the size of the alphabet. The objective is to obtain a mapping from the space of sequences $A^L$ to a feature representation in $\mathbf{R}^N$.

\begin{figure*}

  \centering
    \subfloat[sequence-to-sequence autoencoder]{
      \includegraphics[]{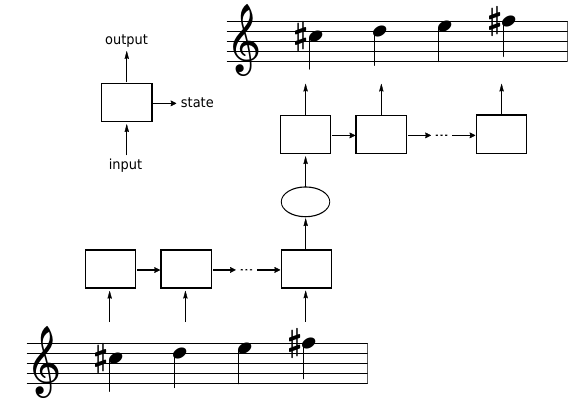}
      \label{fig:ae}       
      }
  \subfloat[transposing sequence-to-sequence architecture]{
    \includegraphics[]{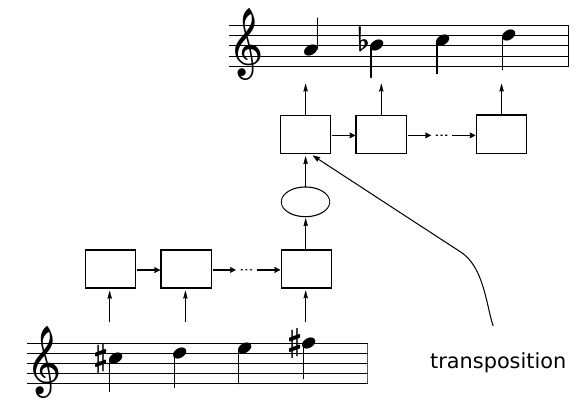}
     \label{fig:aetrans}
    
}

  \caption[AEs]{Seq2seq autoencoder and its generalization. Boxes represent RNN cells. Only the output of the last RNN cell
  is used. The feature representation is displayed as an oval.}
  \label{fig:transpoexample}
\end{figure*}

In order to build a high-level representation of musical sequences in an unsupervised manner, we consider using a
\emph{sequence-to-sequence} model \cite{sutskever2014sequence,DBLP:journals/corr/ChoMGBSB14} (Fig. \ref{fig:ae}) as an autoencoder \cite{Goodfellow-et-al-2016}.
An autoencoder is a Neural Network (NN) parametrized by $\theta$ which is composed of two parts: an encoding NN $\textrm{enc}_\theta$ which usually maps the high-dimensional observation space to a feature representation (or code) of smaller dimensionality and a decoding NN $\textrm{dec}_\theta$ whose aim is to predict the original sequence given its code.

In our case, these neural networks are implemented using Recurrent Neural Network (RNN) \cite{Goodfellow-et-al-2016}. The feature representation $\textrm{enc}_\theta(s)$ for a sequence $s$ is obtained by considering only the output of the RNN on the last time step.

The decoder returns a probability distribution $\pi = (\pi_1, \dots, \pi_L)$ over the sequences in $A^L$ where each $\pi_i = (\pi_{i,1}, \dots, \pi_{i, A})$ is a categorical distribution over $[A]$ ($\pi_{i,a} \geq 0$ and
$\sum_{a=1}^A \pi_{i,a}=1$ for all $i$).

The parameters $\theta$ of the RNN are chosen to minimize the following loss:
\begin{equation}
  \label{eq:4}
  \mathcal{L}(\theta; \mathcal{D}) := \sum_{k=1}^K \sum_{i=1}^N  \textrm{H}\left[s^k_i;  \textrm{dec}_\theta(\textrm{enc}_\theta(s^k))_{i} \right] 
\end{equation}
where
\begin{equation}  
  \textrm{H}\left[s^k_i;  \textrm{dec}_\theta(\textrm{enc}_\theta(s^k))_{i} \right] :=\\
  - \sum_{a=1}^A \delta_a(s^k_i)  \log\left[\textrm{dec}_\theta(\textrm{enc}_\theta(s^k))_{i, a} \right],
\end{equation}
denotes the categorical cross-entropy with $\delta_a$ such that
$\delta_a(x) = 1$ iff. $x = a$ and $0$ otherwise.

Details about our implementation are discussed in Sect.~\ref{sec:impl-deta}.

\subsection{ReLU non-linearity and truncated Spearman rho distance}
\label{sec:relu-non-linearity}
In exposing the rationale behind the permutation-based distance in
Sect.~\ref{sec:pbd}, we put forward the idea that each feature of a
sequence encodes a particular high-level concept. An ordering of these
concepts would then act like a fingerprint for this sequence. However,
in the model described above, the ``most-relevant concepts'' (the
coordinates $x_i$ which have supposedly the greatest impact on the
sequence decoding) are the coordinates with the greatest absolute
value. Coordinates near zero are unlikely to be particularly relevant
but have a great influence on the ordering $\Pi(x)$ of
Eq.~(\ref{eq:2}) and on the distance Eq.~(\ref{eq:1}).

A way to deal with this consists in adding a Rectifier Linear Unit
(ReLU) activation \cite{nair2010rectified} on top of the encoder
RNN. The resulting feature vector will then contain only non-negative
elements with a substantial number of null elements. In doing so, we
impose that the ``most-relevant concepts'' are within the first values
of the permutation $\Pi(x)$. It is then possible to modify the
Spearman rho distance by only considering the $l$ most important
coordinates, which results in the truncated Spearman rho distance of
order $l<N$:
\begin{equation}
  \label{eq:3}
 \rho_l(x, y) = \sqrt{\sum_{i = 1}^l (\Pi(x)_i - \Pi(y)_i)^2}.
\end{equation}

We finally define the corpus-dependent distance $D_\mathcal{D}$ between two sequences $s$ and $s'$ truncated up to $l$ by
\begin{equation}
  \label{eq:6}
  D_\mathcal{D}(s, s'; l) = \rho_l( \textrm{enc}_\theta(s),  \textrm{enc}_\theta(s')),
\end{equation}
where $\textrm{enc}_\theta$ is the encoder RNN (with a ReLU non-linearity on its top-most layer) of a trained sequence-to-sequence autoencoder.

\section{Transformation-invariant distances}
\label{sec:tid}
We now suppose that we have a set of transformations $\mathcal{T}$
that act on sequences. We suppose that this action defines an
\emph{equivalence relation} on $\mathcal{D}$. This means that if there exists
$t \in \mathcal{T}$ such that $s = t.s'$ for sequences $s, s' \in
\mathcal{D}$, then there exists $t' \in \mathcal{T}$ such that $s' =
t'.s$. In such a case, we note $s \sim s'$ and their \emph{equivalence
class} is notated $\mathcal{T}.s$.

A typical example on musical sequences would be
the set of transpositions (see Fig.~\ref{fig:transpoexample}). The
sequence dataset is then split between the different equivalence classes
$\mathcal{T}.s$. Our objective is to devise a distance between sequences that
would be invariant relatively to this set of transformations, i.e. the
distance only depends on the equivalence classes and not on the sequences
themselves.

A simple way to achieve this goal is to directly obtain transformation-invariant
feature representations.  For this, we need to modify the preceding architecture
so that the feature representation (represented as a circle in
Fig.~\ref{fig:ae}) is the same for a sequence and all its transformation.  We
introduce the sequence-to-sequence architecture depicted in
Fig.~\ref{fig:aetrans}. It is a model which takes as an input a sequence
$s \in \mathcal{D}$ and a transformation $t \in \mathcal{T}$ and learns to
predict the transformed sequence $t.s$. In this model, the encoder
$\textrm{enc}_\theta(s)$ is only a function of $s$ while the decoder $\textrm{dec}_\theta(x, t)$ takes as an
input a feature representation $x$ together with the transformation $t$ applied to $s$.

This architecture is trained using the following loss function:
\begin{equation}
  \label{eq:7}
    \mathcal{L}(\theta; \mathcal{D}, \mathcal{T}) := \\ \sum_{t \in \mathcal{T}_k} \sum_{k=1}^K \sum_{i=1}^N  \textrm{H}\left[(t.s^k)_i;  \textrm{dec}_\theta(\textrm{enc}_\theta(s^k), t)_{i} \right], 
  \end{equation}
  where the first sum over $t \in \mathcal{T}_k$  denotes the set of transformations $t \in \mathcal{T}$ such that $t.s^k \in \mathcal{D}$.
 
Note that, in the case of musical transpositions, we can specify how
we want to transpose our input sequence $s$ by two means:
\begin{itemize}
\item \emph{relatively to $s$}, by specifying the interval by which we
  transpose $s$ (relative representation),
\item \emph{independently of $s$}, by specifying, for instance, the
  name of the starting note of $t.s$ (absolute representation).
\end{itemize}

Since we want a feature representation which depends only on
equivalence classes $T.s$ and not on its representatives $s' \in T.s$,
we must use an absolute representation when specifying the
transformations $t \in \mathcal{T}$. Indeed, using a relative
representation would otherwise force the feature representation to
contain absolute information. Plugging this representation in a
rank-based distance would lead to a distance which is not
transformation-invariant.

\begin{figure}
  \centering
      \includegraphics[]{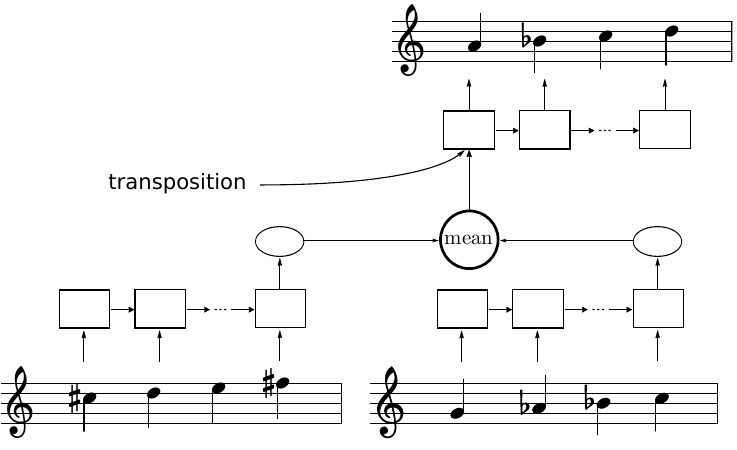}
  \caption{Proposed architecture for computing
    transformation-invariant feature representations. All three
    sequences belong to the same equivalence class.}
  \label{fig:finalArchitecture}
\end{figure}

Even when doing so, the distances we obtain are not fully-invariant:
the feature representation can still contain absolute information
about sequence $s$.  We
propose to overcome this issue by forcing the model to compute
averaged feature representations.
Ideally, using the mean representation
\begin{equation}
  \label{eq:mean}
  \overline{\textrm{enc}}_\theta(\mathcal{T}.s) :=
  \frac{1}{|\mathcal{T}.s|}\sum_{s' \in \mathcal{T}.s} \textrm{enc}_\theta(s')
\end{equation}
for an equivalence class $\mathcal{T}.s$ gives a
transformation-invariant representation, but it is
computationally-expensive.  A simple approximation is to compute this
mean representation using only two sequences belonging to the same
equivalent class.  For two sequences $s, s' \in \mathcal{T}.s$, we
consider the averaged representation
\begin{equation}
  \label{eq:av}
  \overline{\textrm{enc}}_\theta(s, s') :=
  \frac{1}{2}\left(\textrm{enc}_\theta(s) + \textrm{enc}_\theta(s')\right);
\end{equation}
this representation is then passed to the  decoder together
with the absolute transformation representation. This architecture is
represented in Fig.~\ref{fig:finalArchitecture}.

We finally add a $\ell_1$-penalty on the difference between the two
feature representations to encourage the model to make these two
representations equal. The loss function we obtain in this case is then given by

  \begin{equation}
  \label{eq:8}
  \mathcal{L}_\textrm{invariant}(\theta; \mathcal{D}, \mathcal{T}, \lambda) := \\ \sum_{t, t' \in \mathcal{T}_k} \sum_{k=1}^K \sum_{i=1}^N \left( \textrm{H}\left[(t'\!.s^k)_i\ ;  \textrm{dec}_\theta\left(\overline{\textrm{enc}}_\theta\left(s^k, t.s^k\right), t'\right)_{i} \, \right] + \right. \\ \left. \lambda  \ell_1\!\left[\textrm{enc}_\theta(s^k) - \textrm{enc}_\theta(t.s^k)\right] \right),
    \end{equation}

where $\lambda$ is a hyper-parameter controlling the trade-off between accuracy
and enforcing the invariance property.

We thus define a transformation-invariant
$D_{\mathcal{D}, \mathcal{T}}$ distance relative to a corpus
$\mathcal{D}$ and a set of transformations $\mathcal{T}$ as in
Eq.~(\ref{eq:6}), except that the encoder network
$\textrm{enc}_\theta$ is the encoder network of the modified
architecture schematically displayed in
Fig.~\ref{fig:finalArchitecture} and trained with loss  $\mathcal{L}_\textrm{invariant}(\theta; \mathcal{D}, \mathcal{T}, \lambda)$ given by Eq.~\eqref{eq:8}.

This transposition-invariant distance has the advantage that the notion of
invariance is directly encoded into the learned model. Indeed, a simpler way to
produce a transposition-invariant distance would be to apply the corpus-based
distance of Eq.~\eqref{eq:6} to a corpus where all sequences would start with
the same note, say C4. The distance between any two sequences would thus be the
distance between their transposed version starting in C4. However, these
transpositions cannot be implemented effectively: the first note has to be found
in order to know how to transpose (we can have rests or hold symbols, see
Sect.~\ref{sec:impl-deta}). This operation takes some time and it cannot be
easily parallelizable on a GPU. The proposed transposition-invariant distance
thus transfer these costly operations from the evaluation phase to the
data-preprocessing phase.

\section{Experimental results}
\label{sec:res}
We report experiments on the \emph{chorale melodies} from the
\emph{chorale harmonizations by J.S. Bach}. This dataset is obtained
by extracting all soprano parts from the J.S. Bach chorales dataset
included in the music21 \cite{cuthbert2010music21} Python package.

\subsection{Implementation details}
\label{sec:impl-deta}

We choose to encode our data using the melodico-rhythmic encoding proposed in \cite{hadjeres2016deepbach}. More
specifically, time is quantized with sixteenth notes and we use the
\emph{full name} of the notes instead of the traditional MIDI pitch
representation. We introduce two additional tokens in order to handle
rhythm and pitch in a unified fashion: a hold symbol HOLD indicating
that a note is being played but not attacked, and a rest symbol REST.
In this setting, a musical sequence is only an ordered sequence of
tokens drawn from the set of all possible notes \{C3, C\#3, Db3, D3,
D\#3, $\dots$\} together with the HOLD and REST tokens.

The set of transformations $\mathcal{T}$ we choose is the set of all
transpositions. But for a given sequence $s$, we only define as its
equivalence class the set of its transpositions which fits within the original
voice range. In doing so, we do not need the set of transformations to
be a group, but require only that it defines an equivalence relation.

The RNN we use is a 2-layer stacked LSTM \cite{hochreiter1997long} with 512 hidden units each. The ReLU non-linearity is used and the truncation order $l$ of Eq.~(\ref{eq:3}) is set to 256.

\subsection{Nearest neighbor search}
\label{sec:NN}
\begin{figure}
  \centering
  \subfloat[Corpus-dependent distance $D_\mathcal{D}$ constructed using Spearman's rho distance]{
      \includegraphics[width=0.53\textwidth]{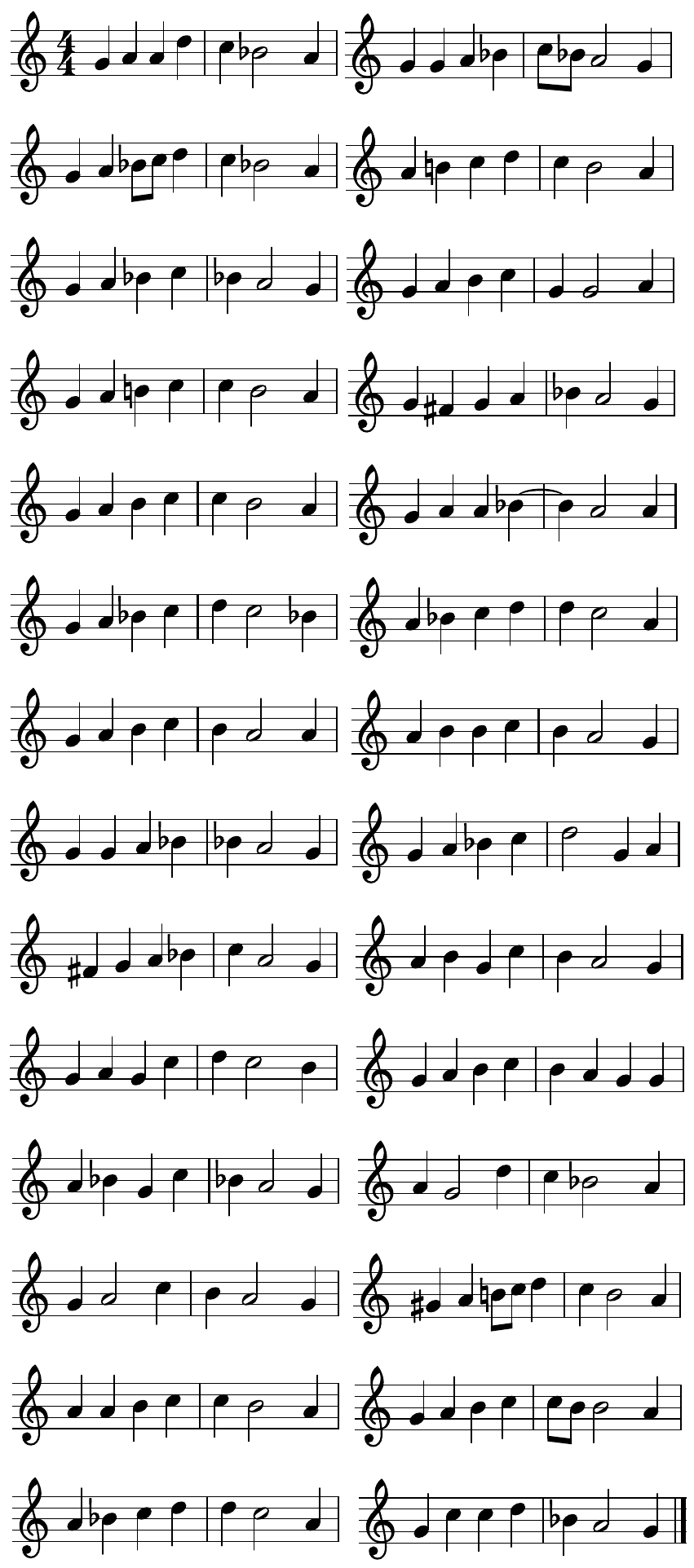}
  \label{fig:nearest:spearman}
      }
      \subfloat[Edit distance]{
\includegraphics[width=0.355\textwidth,clip=true,trim=0 -5 0 0]{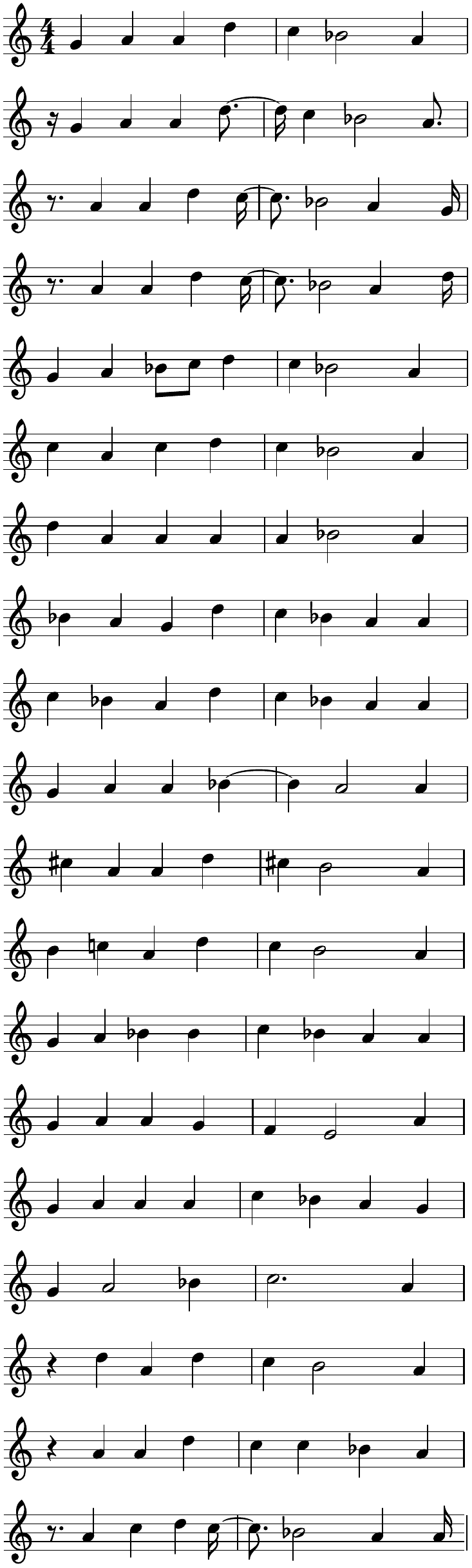}
        \label{fig:nearest:edit}

}

  \caption[AEs]{A target melody and its nearest neighbors according to different distances on sequences. Duplicates have been removed.}
  \label{}
\end{figure}

We empirically demonstrate the efficiency of these distances by displaying examples of nearest neighbor requests.
Fig.~\ref{fig:nearest:spearman} shows examples of melodies which are ``close'' according to the corpus-dependent distance $D_\mathcal{D}$. All the results that we display are obtained using Spearman's rho distance, but we obtain similar results by replacing it with Kendall's tau similarity measure.

The nearest neighbors query  of Fig.~\ref{fig:nearest:spearman} reveals interesting behaviors of the corpus-dependent distance $D_\mathcal{D}$. The overall shape of the target melody (ascending then descending) is present in every neighboring sequence under various aspects.  There are interesting rhythmic variations and also key changes. But, what is the most striking here, is that some other characteristic elements of the target sequence are also taken into account. For instance, the importance of the ascending leap on the fourth beat is present in many sequences. Another such example is the note repetition at beats 2 and 3 which also occurs in some neighboring sequences. We believe that the last sequence in the example displayed in Fig.~\ref{fig:nearest:spearman} is a good illustration of how characteristic elements are captured with our distance. Even if there are only two notes in common with the target sequence, we still have an impression of proximity between both. This may be due to the following facts
\begin{itemize}
\item  the most important notes of the target sequence (the highest and the lowest) are replicated,
  
\item the key and the rhythm are identical,
  
\item there is an ascending fourth in both (on beat 2 for the neighboring sequence instead of on beat 4 for the target sequence),
  
\item there is a note repetition on beats 2 and 3,
  
\item they both conclude by a descending conjunct movement.
\end{itemize}

This has to be compared with the nearest neighbors returned using the edit
distance on the same target sequence. This is displayed in
Fig.~\ref{fig:nearest:edit}. This presents some unwanted behaviors. For
instance, sequences identical to the target sequence but with a sixteenth note
offset are considered to be almost identical. However, from a musical point of
view, the importance of the difference between notes on and notes off the beat
is crucial. In the other cases, the edit only manages to find sequences
containing common notes (and played at the exact same time) with the target
sequence. Since the HOLD symbol is seen as note like any other one, we can see
that replacing a half note by a quarter note has a cost of only one,
independently of the chosen note. The importance here is not on the melodic
contours nor on an intuitive perception of similarity.

The behavior of the edit
distance seems more erratic and less predictable. It is also ``less smooth''
than our proposed distances (see Fig.~\ref{fig:transpoexample}) since the edit
distance can only take a finite (and smaller) number of values. The important
difference is that the number of possible values returned by the edit distance
depends on the sequences size, while it depends on the feature vector size in
the case of corpus-dependent distances.

\subsection{Invariance by transposition}
In this section, we check to which extent the distance $D_{\mathcal{D}, \mathcal{T}}$ is invariant under the set of transformations $\mathcal{T}$. In Fig.~\ref{fig:transpoexample}, we plot the distance $D_{\mathcal{D}, \mathcal{T}}(s, s')$ when $s$ and $s'$ are in the same equivalence class under $\mathcal{T}$ and when they are not.

\begin{figure*}
  \centering
    \subfloat[Using Spearman's rho rank distance]{
      \includegraphics[width= 0.46\textwidth]{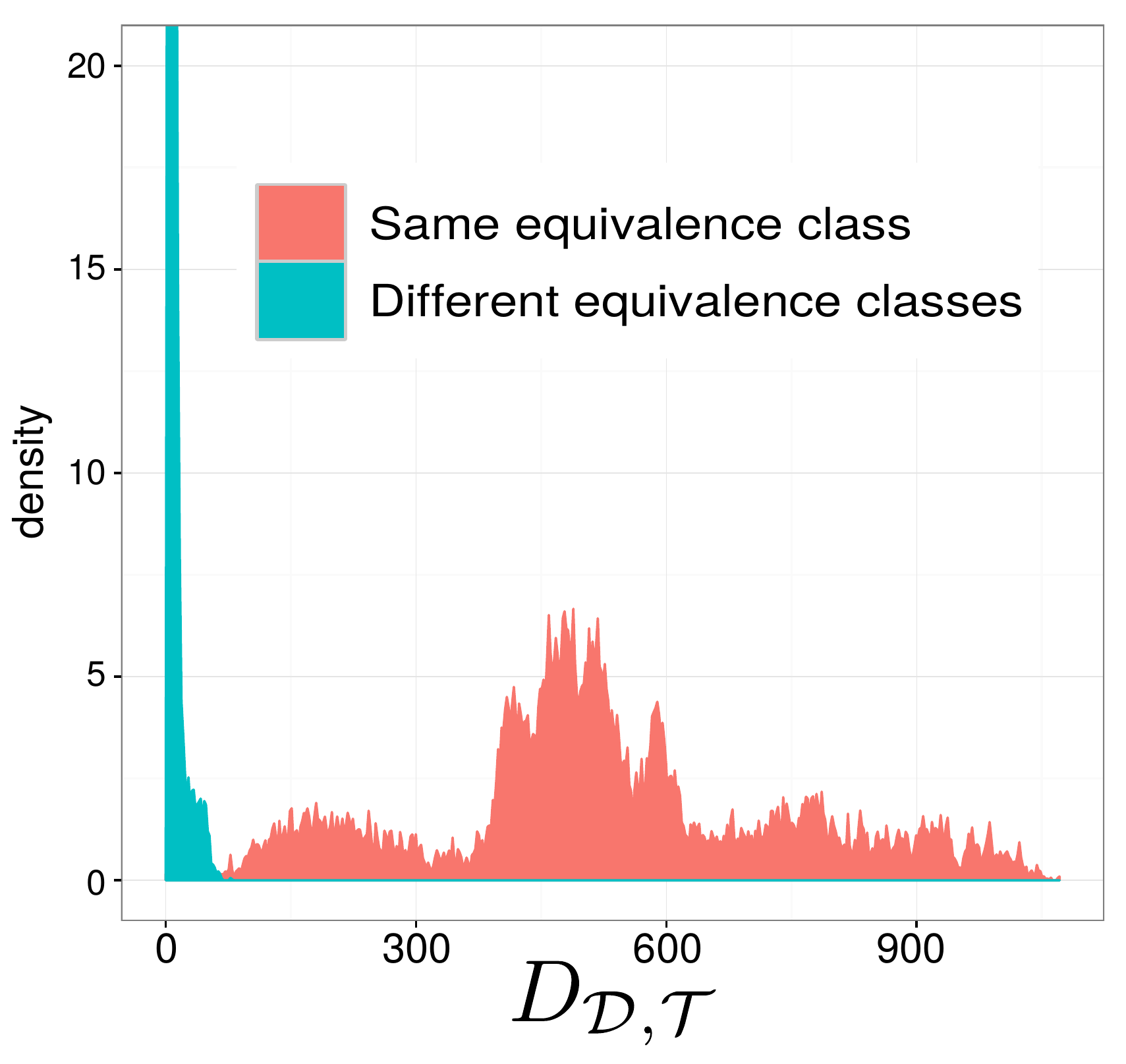}
      \label{fig:stats:spearman}       
      }
  \subfloat[Using Kendall's tau rank distance]{
    \includegraphics[width= 0.46\textwidth]{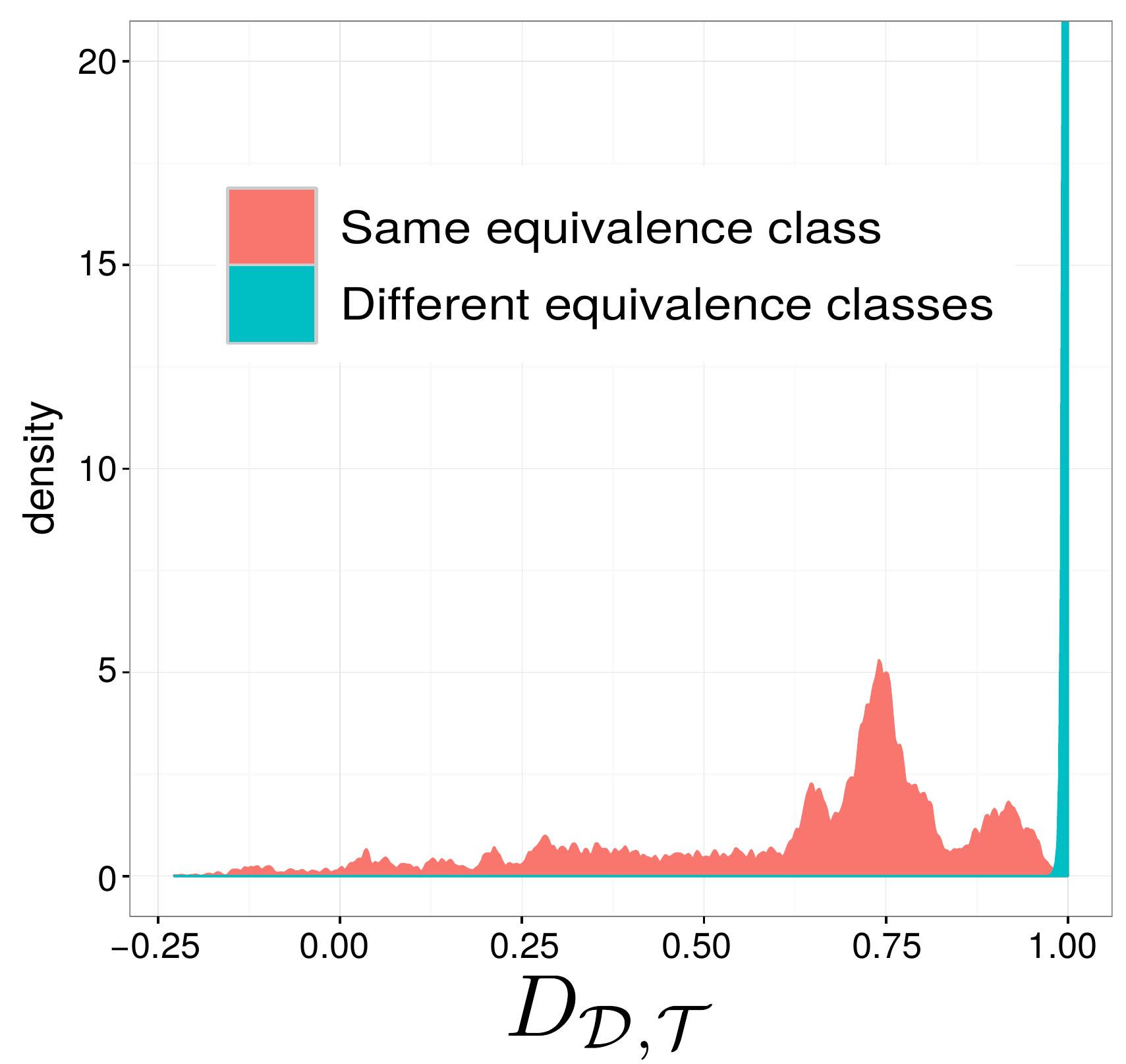}
     \label{fig:stats:kendall}
    
}

  \caption[AEs]{Density estimations of  distances $D_{\mathcal{D},
         \mathcal{T}}(s, s')$ between two sequences $s$ and $s'$ belonging to
       the same or to different equivalence classes. The two proposed rank distances are compared.
      Our architecture allows to
       capture the invariance by transposition for musical sequences (peaked
       blue). The images are truncated for clarity.}
  \label{fig:transpoexample}
\end{figure*}

Two things are two be noted in this example. Firstly, the obvious difference between the two densities and the sharp peak when the sequences are in the same equivalence class show that our distance indeed captures the invariance by transposition on musical sequences. Secondly, the  behavior of the density of the distance between two randomly-chosen sequences is interesting: it is multimodal and widespread. This distance can take numerous different values and can then have more discriminative power. We have experimentally seen in Sect.~\ref{sec:NN} that the corpus-dependent distance is able to capture high-level musical concepts. It is the same for its transposition-invariant counterpart.

\begin{figure*}
  \centering
  \includegraphics[scale=0.8]{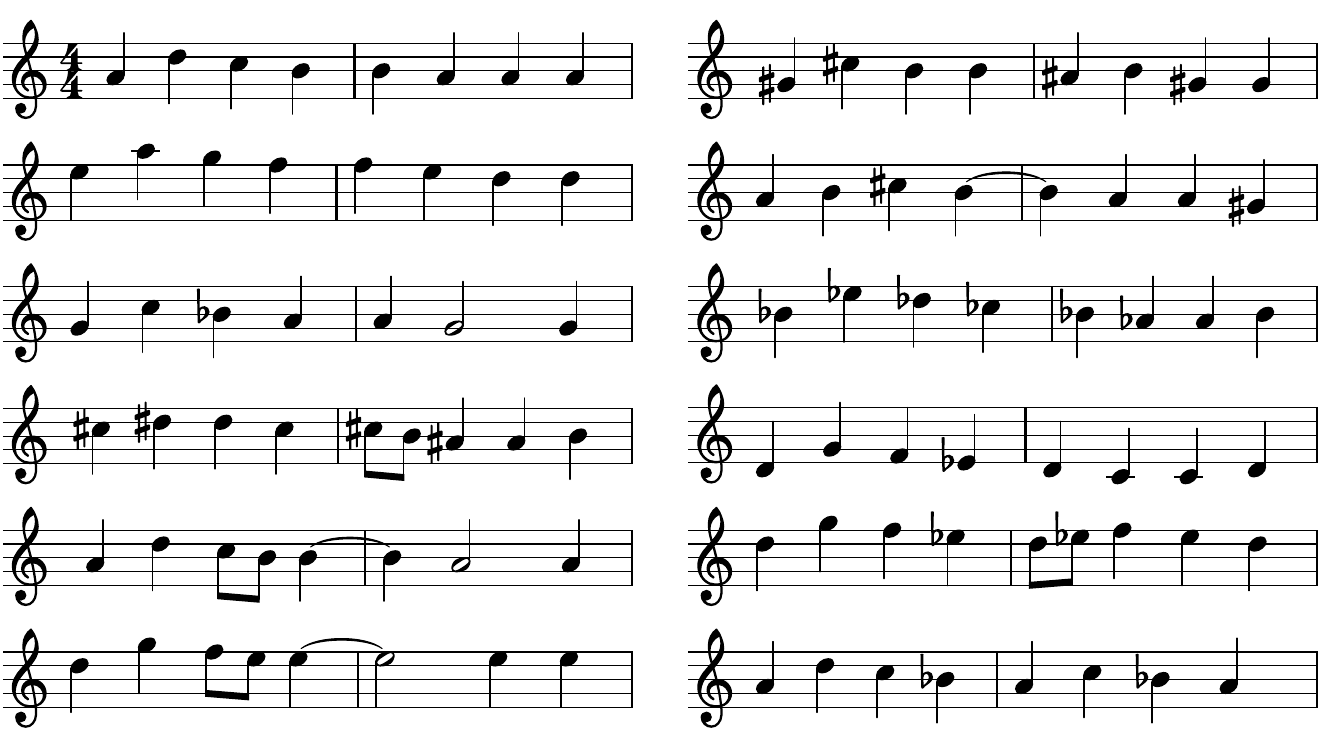}
    \caption{A target melody and its nearest neighbors among 20000 sequences
      randomly drawn from all possible subsequences using the
      transposition-invariant distance  $D_{\mathcal{D}, \mathcal{T}}$ based on the Spearman
      rank-based distance.}
  \label{fig:nnInv}
\end{figure*}

To show this, we do not replicate the results about the nearest neighbor search of Fig.~\ref{fig:nearest:spearman}  since it only returns exact transpositions of the target sequence. Instead, we only make a nearest neighbor search on a small subset of the J.S. Bach chorales using  $D_{\mathcal{D}, \mathcal{T}}$. The result is displayed in Fig.~\ref{fig:nnInv}.

The same analysis as the one conducted for Fig.~\ref{fig:nearest:spearman} can be made, except that it is now irrespective of the transposition of the neighboring sequences. We conclude that this transposition-invariant distance allows to detect characteristic patterns and musical motives independently of the key they are in.

\section{Conclusion}
\label{sec:conclusion}
We proposed a novel framework to build distances learned from musical
corpora. Because they take into account the style to which musical sequences
belong, these learned distances are not subject to the usual problems
encountered using the edit-distance generalizations: they are less dependent on
the input encoding while being more satisfactory from a perceptive point of
view. 
Indeed, using neural-network-learned features instead of handcrafted features
allows to define a natural notion of proximity between sequences which is rooted on an
objective and non-biased a priori basis. The choice of the rank-based distance applied on
the feature representations does not seem to influence much the final distance
behavior.
This framework can be modified so that the
distances we obtain are invariant with respect to a given a set of
transformations. This is made possible by the fruitful combination between a
quasi-invariant feature representation learned from a regularized
sequence-to-sequence architecture and a rank-based distance. Since the feature
representations are not necessarily equal for sequences within the same
equivalence class, the usage of a rank-based distance over these representations
helps to make the distance over sequences invariant.

Future work will aim at improving the proposed method by taking into account multiple hidden layers in the rank-based distance.
On a more practical side, we will also aim at using this distance on music generation tasks in
order to design algorithms capable of generating highly-structured melodies.

\section*{Acknowledgments}
\label{sec:acknowledgments}
First author is funded by a PhD scholarship (AMX) from \'{E}cole Polytechnique.

\bibliographystyle{plain}
\bibliography{variations}

\end{document}